# HEALTH AND MEDICINE

# Effective drug combination for *Caenorhabditis elegans* nematodes discovered by output-driven feedback system control technique

Xianting Ding,[1]*[†] Zach Njus,[2][†] Taejoon Kong,[2] Wenqiong Su,[1] Chih-Ming Ho,[3] Santosh Pandey[2]*

Infections from parasitic nematodes (or roundworms) contribute to a significant disease burden and productivity losses for humans and livestock. The limited number of anthelmintics (or antinematode drugs) available today to treat these infections are rapidly losing their efficacy as multidrug resistance in parasites becomes a global health challenge. We propose an engineering approach to discover an anthelmintic drug combination that is more potent at killing wild-type *Caenorhabditis elegans* worms than four individual drugs. In the experiment, freely swimming single worms are enclosed in microfluidic drug environments to assess the centroid velocity and track curvature of worm movements. After analyzing the behavioral data in every iteration, the feedback system control (FSC) scheme is used to predict new drug combinations to test. Through a differential evolutionary search, the winning drug combination is reached that produces minimal centroid velocity and high track curvature, while requiring each drug in less than their $EC_{50}$ concentrations. The FSC approach is model-less and does not need any information on the drug pharmacology, signaling pathways, or animal biology. Toward combating multidrug resistance, the method presented here is applicable to the discovery of new potent combinations of available anthelmintics on *C. elegans*, parasitic nematodes, and other small model organisms.

## INTRODUCTION

Resistance of pathogens to pharmaceutical drugs is a widely acknowledged problem in medicine (*1*–*7*). Parasitic nematodes are a classic example of multidrug resistance, because these animals have evolved resistance to chemotherapy through successive generations (*8*–*13*). Neglected tropical diseases arising from parasitic nematode infections contribute to extensive disease burden and socioeconomic losses, particularly in the underdeveloped countries of the world (*14*–*18*). There are a number of contributing factors for the emergence of multidrug resistance in parasitic nematodes. For instance, there are a limited number of drugs approved by the World Health Organization (WHO), the cost and time for new drug discovery are astounding, existing chemotherapy options require multiple treatments and doctor visits, large-scale deworming efforts on millions of infected people each year is a daunting task, and there is a large geographic variation in the nature and dynamics of parasitic infection (*19*–*26*).

The general classes of compounds, called anthelmintics, have been developed to treat infections by parasitic nematodes (*13*, *21*, *27*). One class of anthelmintics approved for use by WHO is the nicotinic acetylcholine receptor (nAChR) agonist, which can be further divided into two targeted receptor subtypes: L-type and N-type (*27*). Common examples of L-type nAChR agonists are levamisole and pyrantel, whereas an example of an N-type AChR agonist is methyridine (*21*, *27*, *28*). Tribendimidine, a relatively understudied drug, is the only anthelmintic that is undergoing phase 4 clinical testing in China and shows promise against a broad range of parasitic nematode infections (*4*, *5*). With the repertoire of present-day anthelmintics, it would be difficult to imagine that parasitic nematodes and their associated infections continue to be a pressing issue to the society. However, challenges exist in both the frontiers of drug discovery and testing. Besides higher throughput compound screening assays in the laboratory, we need methods to monitor and quantify resistance of nematodes in the field, cost-efficient ways to translate basic research to field testing, and models to predict interactions between multiple drugs when used in rotation or in combination (*18*, *29*–*33*).

Because of the apparent challenges in conducting large-scale drug tests on parasitic nematodes, researchers have often resorted to the free-living *Caenorhabditis elegans* to carry out experiments to determine novel drug targets or molecular modes of action and resistance (*33*, *34*). Among the many benefits, *C. elegans* is easy to culture on agarose plates, has a short life span, is amenable to genetic and mutagenesis tools, and exhibits genetic tractability through generations (*35*, *36*). Technologies, such as optofluidics and microfluidics, have been developed around *C. elegans* to test how drug compounds affect their viability and movement, detect effects of toxins on their behavior and life cycle, and decipher the role of individual genes in producing a specific behavioral response (*36*–*40*). High-resolution imaging tools with sophisticated worm tracking software have been realized to automatically generate data on behavioral phenotype, thus minimizing the role of manual screening (*41*–*43*). Technological advancements have definitely improved the pace of drug testing with the ability to screen thousands of small-molecule compounds for effectiveness in *C. elegans* (*44*, *45*). However, the process of drug discovery still relies on the conventional approach of searching for suitable molecular targets and understanding the modes of action of a new compound (*34*). In an attempt to address this bottleneck, we questioned whether a control engineering approach could be used to speed up the process of drug discovery wherein an algorithmic search helps us achieve the desired drug effectiveness against parasitic nematode infections (*46*).

Here, we show that a specific combination of four anthelmintics (namely, levamisole, pyrantel, methyridine, and tribendimidine) is more effective at killing wild-type *C. elegans* than each of the four anthelmintics used separately. A schematic representation of our experiment is illustrated in Fig. 1. Active worms are placed in microfluidic chambers where they are free to swim around in a drug environment. Their movement behavior is recorded in response to various concentrations of the four anthelmintics. The dose-response data for individual drugs are fed into a feedback system control (FSC) optimization scheme (46–48) that suggests new combinations of the four drugs to test during each iteration. Using a custom worm tracking program, we record the movement coordinates of single worms at real time. We found that two movement parameters, average centroid velocity and track curvature, adequately represent how certain drug combination affects the worm. Then, through multiple iterations of testing drug combinations, the FSC technique is able to produce the winning cocktail that is more potent but uses less than the median effective concentration ($EC_{50}$) values of the four anthelmintics.

[1]State Key Laboratory of Oncogenes and Related Genes, Institute for Personalized Medicine, School of Biomedical Engineering, Shanghai Jiao Tong University, Shanghai, China. [2]Department of Electrical and Computer Engineering, Iowa State University, Ames, IA 50011, USA. [3]Department of Mechanical and Aerospace Engineering, University of California, Los Angeles, Los Angeles, CA 90095, USA.
*Corresponding authors. Email: dingxianting@sjtu.edu.cn (X.D.); pandey@iastate.edu (S.P.)
[†]These authors contributed equally to this work.

## RESULTS
### Dose response of individual drugs

To start with, four anthelmintic drugs are chosen that are known to act on the *C. elegans* worms (49–53). Different concentrations of each drug are prepared in M9 buffer. For each concentration, the microfluidic chamber is filled with the drug solution and worm behavior is recorded for 600 s at one frame per second. This time duration is sufficiently long to observe the drug-induced degradation in the worm's physical ability (54–57). The average velocity of the worm's centroid in an experimental run is calculated and normalized to that obtained in control experiments with M9 buffer to obtain a percentage response (that is, dose response). Each drug concentration is tested on at least three separate chips with one worm per chamber and three chambers per chip. Figure 2 shows the dose response of *C. elegans* to the four drugs from which the $EC_{50}$ value is calculated (that is, drug concentration where the percentage dose response is 50%). The $EC_{50}$ values obtained from our experiments are as follows: levamisole, 2.23 μM; pyrantel, 107.5 μM; tribendimidine, 4.68 μM; and methyridine, 1672 μM. From the measured $EC_{50}$ values, it appears that levamisole is the most potent drug, whereas methyridine is the least potent one.

### FSC implementation

After obtaining the dose response of the four drugs, we proceed to use the FSC technique to discover a new combination that is more potent than the $EC_{50}$ concentrations of any individual drug. In other words, we want to check the possibility of creating a concoction using the four drugs at reduced concentrations but with higher potency or greater efficacy on worm activity (that is, lower average centroid velocity, typically <15 μm/s). This will allow us to use lower drug dosages, thereby reducing the associated costs. To use the FSC technique, we choose six concentration keys for each of the four drugs. A concentration key refers to a representative, integer value assigned for an actual concentration of a drug. The FSC technique does not use actual drug concentrations as input but rather requires concentration keys to determine the drug combinations for the subsequent steps. The concentration keys are chosen as follows: The highest concentration key is the $EC_{50}$ value of the drug. The next lower concentration key corresponds to half of the present concentration. The concentration keys and their respective drug concentrations are listed with specific color codes for each drug (Fig. 3, A and B).

For the first iteration, all the concentration keys are inputted into the FSC program, which then outputs eight drug combinations to test (that is, hereby called P1 to P4 and T1 to T4). Figure 3C illustrates the process flow where a drug combination is represented by a set of four unique numbers denoting the concentration keys of the drugs used according to Fig. 3 (A and B). As an example, P1 in iteration 1 is represented as "4524", which indicates the following concoction: levamisole, 1 μM; tribendimidine, 4 μM; pyrantel, 12.5 μM; and methyridine, 800 μM. Each drug combination is prepared, filled in the microfluidic chambers, and tested on single L4-stage *C. elegans*. Each drug combination is tested on at least three to five different chips, with each chip having three separate chambers with one worm per chamber. From the recorded videos, the average centroid velocity is obtained and normalized to values from the control experiments, and the percentage dose response (or performance parameter) is calculated for all the eight drug combinations.

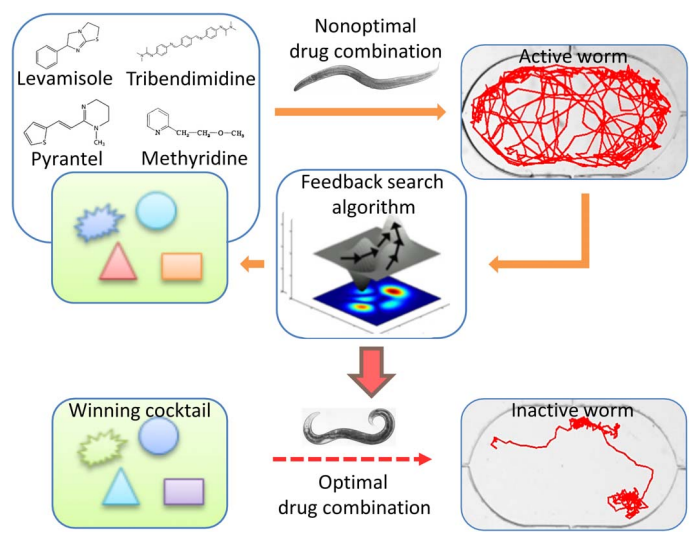

**Fig. 1. Schematic representation of the experiment.** The effect of the four anthelmintics (namely, levamisole, pyrantel, methyridine, and tribendimidine) is tested on active, wild-type *C. elegans*. The dose-response data are fed into an FSC scheme that suggests new combinations of the four anthelmintics to test for worm survivability. Through multiple iterations of trying nonoptimal combinations and using a directed evolutionary search, the winning cocktail is eventually reached that makes the worm inactive, as judged by the values of average centroid velocity and track curvature.

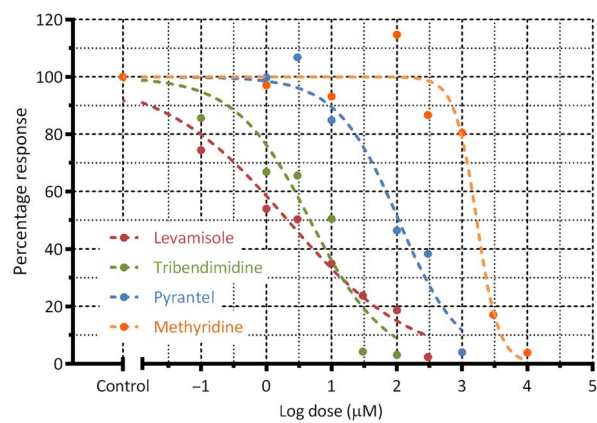

**Fig. 2. Dose response of the four anthelmintics is shown.** Here, the percentage response is obtained by calculating the average centroid velocity of a worm population at a certain drug concentration and normalizing it to the velocity from the control experiments with M9 buffer. The $EC_{50}$ (that is, drug concentration where the percentage response is 50%) values obtained from our microfluidic device are as follows: levamisole, 2.23 μM; pyrantel, 107.5 μM; tribendimidine, 4.68 μM; and methyridine, 1672 μM. Each drug concentration is tested on at least three separate chips with one worm per chamber and three chambers per chip.

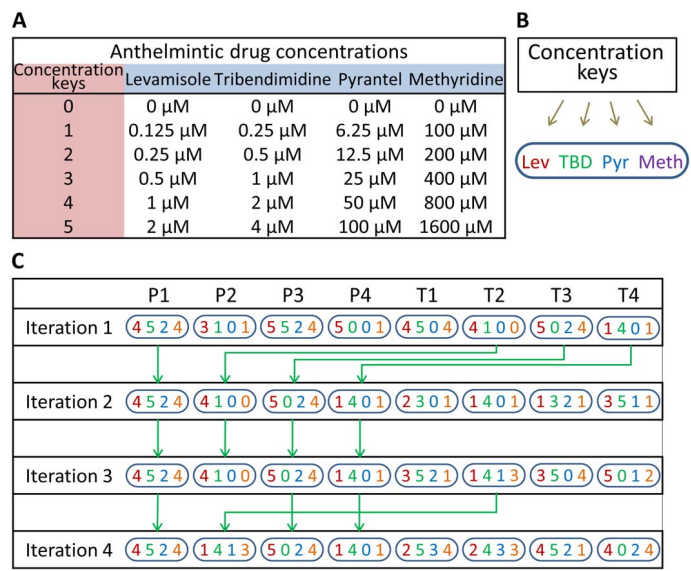

Fig. 3. **Drug concentration keys and FSC iterative process flow.** (**A**) The concentration keys (0 to 5) are assigned to six individual concentrations of each anthelmintic. The highest concentration key (that is, key 5) is chosen close to the $EC_{50}$ values, thus limiting our search to a winning combination having no more than the $EC_{50}$ value of each compound. (**B**) Distinct color codes are chosen for each drug for visual representation. (**C**) In each iteration, the eight combinations are grouped into P group (P1 to P4) and T group (T1 to T4) candidates. The first iteration tests eight combinations. The second iteration retains the four best combinations from the first iteration as the P group and suggested four new combinations as the T group. This procedure is repeated in each iteration until the winning combination is reached (T3 in the fourth round here).

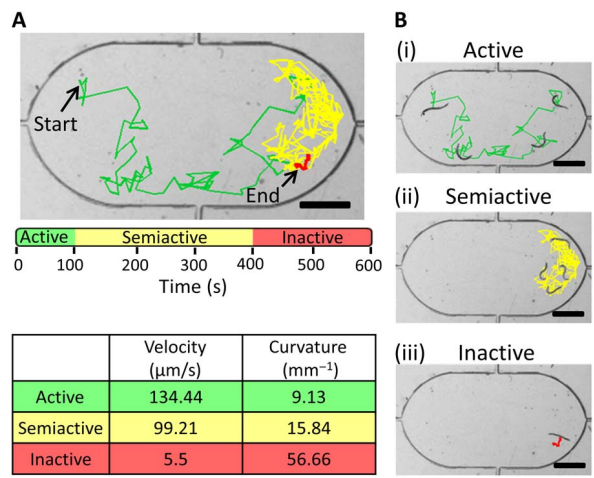

Fig. 4. **Microfluidic device and worm tracking.** (**A**) Movement of a wild-type *C. elegans* worm is tracked through a 600-s video where it shows distinct periods of being active, semiactive, and inactive. An active worm has relatively high velocity and smaller curvature, whereas an inactive worm displays minimal velocity and larger curvature. (**B**) Tracks of the same worm [as in (A)] are shown during the active, semiactive, and inactive periods (i to iii). The active worm has relatively straight tracks with a smaller curvature, whereas the inactive worm shows considerable struggle with minimal displacement and a larger curvature. Scale bars, 400 μm.

For the second iteration, the performance of the P and T groups from the first iteration is compared (that is, P1 versus T1 and P2 versus T2, etc.). The four better-performing drug combinations are retained as P1 to P4, and the four lower-performing drug combinations are discarded. In addition, the FSC program suggests four new combinations as T1 to T4 that are also tested along with the P group in the second iteration. The experimental process is repeated where the eight drug combinations are prepared and tested on *C. elegans* within microfluidic chambers. The performance of the eight combinations is evaluated to determine whether the effective or winning drug combination is reached. Otherwise, the above procedure is repeated for successive iterations until the winning drug combination is achieved. Figure 3C shows the P and T group drug combinations for every iteration. In our case, the winning combination was reached after the fourth iteration.

### Average centroid velocity and track curvature
From our initial observations, it was evident that worms in the drug environment showed three distinct categories of movement activity (that is, active, semiactive, and inactive). In a potent drug solution, worms initially show unrestricted swimming patterns (active period) that slowly change to restricted swimming with more struggle (semiactive period) and eventually to an almost immobile posture (inactive period). Figure 4A shows the superimposed tracks of a single worm in a potent drug solution through 600 s of recording with its three periods of activity which are color coded for visual ease. In this example, the active worm exhibits relatively high velocity (134.44 μm/s) and smaller track curvature (9.13 $mm^{-1}$), the semiactive worm shows medium velocity (99.21 μm/s) and track curvature (15.84 $mm^{-1}$), whereas the inactive worm displays minimal velocity (5.5 μm/s) and larger track curvature (56.66 $mm^{-1}$). Figure 4B shows the tracks of the same worm plotted separately during the active, semiactive, and inactive periods. An active worm moves in a nearly straight track leading to a smaller value of track curvature, whereas an inactive worm struggles considerably with minimal body displacement yielding a larger value of track curvature.

The average centroid velocity and track curvature obtained from the four rounds of iterations are plotted in Fig. 5 (A and B), respectively. Each control and experimental drug combination is tested on at least three to five different chips, with each chip having three separate chambers with one worm per chamber. The data plotted are only from the final 120 s of each video recording that lasted 600 s, although the data averaged over the entire 600 s showed similar trends (data not shown). It is interesting to compare the velocity and curvature data alongside each other. The first iteration that proposed eight drug combinations led to four desirable combinations (that is, P1, P2, T3, and T4) for the second iteration. Unfortunately, the T group performed worse than the P group in the second iteration. The third iteration found four desirable combinations (that is, P1, P3, P4, and T2) that were carried over to the fourth iteration. The T group performed much better than the P group in the fourth iteration, particularly T3 that is chosen as the effective or winning drug combination.

We found that both the centroid velocity and track curvature data are needed to adequately represent the movement behavior of the worms. These data can either be instantaneous or averaged over a certain time period. Figure 6 shows another way to view the data sets. Here, we plot the instantaneous velocities and curvature values for eight representative drug combinations to show the general trends we observed. Some combinations have very high velocity (>100 μm/s) and low curvature (<25 $mm^{-1}$), indicating that the drugs have a negligible effect on worm movement within the observed time period (Fig. 6, D to F). Other combinations may have medium velocity (50 to 100 μm/s) and curvature (25 to 50 mm$^{-1}$), suggesting that the drugs are affecting the worms but better alternatives are still possible. The winning combination has minimal velocity (<10 μm/s) and large curvature (>50 mm$^{-1}$) throughout the 600 s (Fig. 6A). We saw that the winning drug combination started paralyzing the worm within the first minute itself, and the decreased movement was sustained through the 600 s of recording. This suggested that the length of each experiment could be curtailed to within 2 min, and the movement parameters (that is, instantaneous velocity and curvature data) could be used as a rapid indicator of potency.

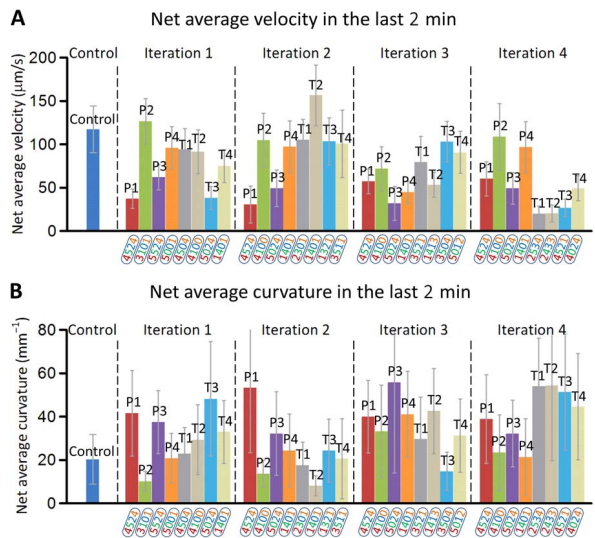

Fig. 5. **Average centroid velocity and track curvature.** (**A**) The centroid velocity averaged over all the worms for a given drug combination is plotted through all the four iterations. The centroid velocity of a single worm is averaged over the last 120s of the 600-s duration of the worm exposure. (**B**) Similarly, the track curvature averaged over all the worms for a given drug combination is depicted through the four iterations. Centroid velocity and curvature of worms in control experiments using M9 buffer are also shown. The winning combination was chosen as T3 in the fourth iteration that produced minimal centroid velocity and a larger curvature. Each drug combination is tested on at least three to five different chips. Each chip has three separate chambers with one worm per chamber. Bars are SEM.

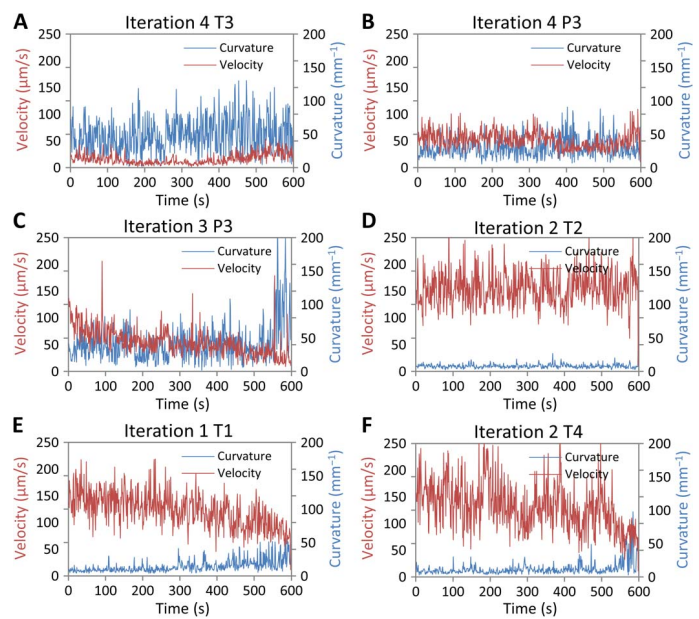

Fig. 6. **Instantaneous centroid velocity and track curvature are plotted for the entire 600 s of drug exposure.** The data at every second are obtained by averaging the velocity or curvature values for all the worms exposed to a certain drug combination. (**A**) The winning combination, T3 of the fourth iteration, has minimal velocity (<10 μm/s) and large curvature (>50 mm$^{-1}$) throughout the 600 s. (**B** and **C**) These combinations show medium velocity (50 to 100 μm/s) and curvature (25 to 50 mm$^{-1}$), suggesting that the compounds affect the worms, but better alternatives are still possible. (**D** to **F**) These combinations exhibit high velocity (>100 μm/s) and low curvature (<25 mm$^{-1}$), indicating that the drugs have a negligible effect on worm movement within the observed time period.

The relationship between centroid velocity and track curvature as measured in the FSC technique can be illustrated as a two-dimensional plot (Fig. 7). The color codes in the plot represent the worm in the three periods of activity: active, semiactive, and inactive. In this depiction, we are interested in potent combinations that make the worm inactive with minimal velocity and larger curvature. In iteration 1, two combinations (P1 and T3) induce inactivity in the worm, P3 makes the worm semiactive, and the remaining combinations do not seem to affect the movement of the active worm (Fig. 7A). Similarly, iteration 2 yields P1 as a good candidate, P3 makes the worm semiactive, whereas the rest of the combinations do not show an effect on the movement of the active worm (Fig. 7B). Iteration 3 produces no combination that makes the worm inactive; however, four combinations (P1, P3, P4, and T2) make the worm semiactive (Fig. 7C). Iteration 4 yields two combinations (T1 and T3) that induce inactivity in the worm, we chose T3 as the winning drug combination that has the minimum centroid velocity (Fig. 7D). The winning combination (T3 of fourth iteration) consists of 1 μM levamisole, 4 μM tribendimidine, 12.5 μM pyrantel, and 100 μM methyridine.

## DISCUSSION

We showed that an optimal combination of four anthelmintics can be found that uses much less than the EC$_{50}$ concentrations of each drug but is more potent than any of them. The experimental data, gathered from the movement dynamics of a single *C. elegans* enclosed in microfluidic chambers, were fed into an FSC optimization scheme. Through a differential evolution search and feedback calculation process, the algorithm produced sets of drug combinations to test in each iteration that eventually led to the winning combination. The interplay of the three technologies (that is, microfluidics, worm image tracking, and FSC search) was crucial to the successful demonstration of the winning drug combination. Microfluidic devices enabled us to record the behavior of freely swimming worms in enclosed chambers. The worm tracking program allowed us to process the recorded videos and extract the average centroid velocity and track curvature for every experiment. The FSC technique provided the means to predict new combinations and optimize our search with every iteration. In our knowledge, this is the first use of microfluidics to discover effective drug combinations on a small-animal model organism through a systematic FSC search. Furthermore, this work validates the workability of the FSC search technique on *C. elegans* model organisms beyond its current usage to cell-based systems (for example, in cancer cell viability, viral infection, stem cell maintenance and differentiation, and endothelial cell viability) (46–48).

Our rationale for the choice of the four drugs was based on our previous experiments (37). From analyzing the behavioral patterns of worm paralysis in individual drugs, we found that tribendimidine was more potent than levamisole or pyrantel in causing irreversible paralysis, methyridine acted on a different subtype of nAChR receptors than the other three anthelmintics, and pyrantel was more effective than levamisole, although they both had similar modes of action (*17*). Thus, on the basis of differences in phenotype and paralysis observed earlier, we started with these four anthelmintics for our FSC-guided search for optimal drug combinations. With every possibility, there may be other drug compounds that lead to concoctions with better potency than that reported here.

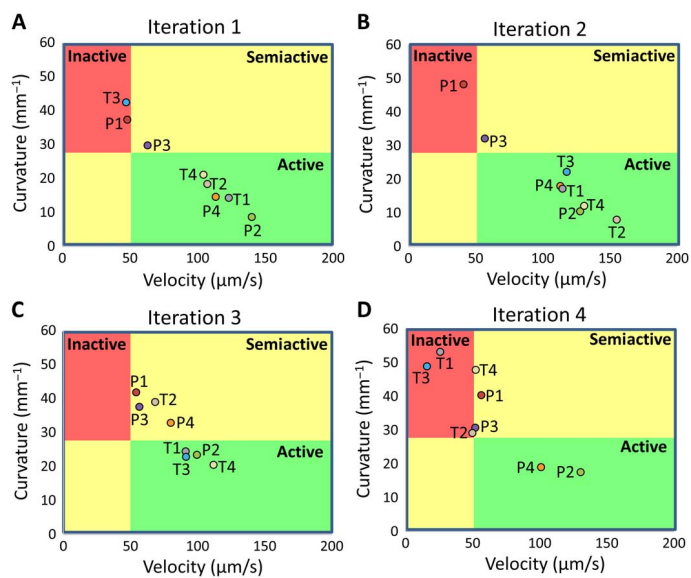

**Fig. 7. Two-dimensional plot of track curvature versus centroid velocity.** In the drug environment, the worm is classified as active, semiactive, or inactive as demonstrated in Fig. 4, and the plot is accordingly segregated into the three areas of activity. We are primarily interested in combinations that induce inactivity in the worm. (**A**) In iteration 1, two combinations (that is, P1 and T3) are able to make the worm inactive. Besides in P3, the active worm does not succumb to any other combination and is classified as being active. (**B**) In iteration 2, only P1 makes the worm inactive. The rest of the combinations do not seem to affect the active worm. (**C**) In iteration 3, the worm is in a semiactive or active state for all combinations. (**D**) In iteration 4, three combinations (T1, T2, and T3) are able to make the worm inactive, and the winning combination is chosen as T3.

The FSC search technique relies on the user's judgment to choose the most suitable drug library (for example, preferably with distinct mechanisms, molecular targets, or measured effects) for the desired outcome. However, a major challenge in choosing an appropriate drug library lies in the fact that the number of drug combinations increases exponentially with increasing number of drugs or doses. As an example, assume that the drug library consists of "n" drugs with "m" concentrations of each drug to be tested. Each concentration is further tested at least three times for consistency. This would involve conducting (m^n) × 3 experiments. Here, we used n = 4 drugs, m = 6 concentration keys, and 9 to 12 worms for each concentration, which gave a total of at least 10,000 single worm experiments to run. If we chose more number of drugs, the number of experiments would be overwhelming with conventional methods. Therefore, the choice of the initial drug library could involve drug candidates that cover multiple signaling pathways and potentially contribute to the end point system readouts. However, if the initial drug library is too large, FSC takes longer iterations to downselect those nonsignificant drugs.

The FSC search technique is a powerful tool toward drug discovery. It is a model-less approach that does not require information about the drug's pharmokinetics or the genetics/metabolomics of the animal under study (as opposed to models based on big data and machine learning). No information is needed about how drug signaling pathways work or how multiple drugs interact in combinations. This model-less approach is advantageous in scenarios where data sets are limited in size and experiments are time-consuming. Here, the assumption is that we do not need a complete model of the complex biosystem (that is, *C. elegans*) to produce a desired output from input variables. The FSC technique uses the system's output (that is, movement data) as the feedback criteria to optimize the input values (that is, individual drug concentrations), eventually reaching the desired output for the winning drug combination.

We demonstrated a specific combination of four anthelmintics that uses drug dosages lower than the individual $EC_{50}$ values of each drug but is still much more potent on *C. elegans* than any of these drugs administered individually. The experiment involves recording the movement of freely swimming single *C. elegans* in microfluidic chambers for 600 s, along with a tracking program to automatically extract the average centroid velocity and track curvature. The $EC_{50}$ values measured are 2.23 μM for levamisole, 107.5 μM for pyrantel, 4.68 μM for tribendimidine, and 1672 μM for methyridine. Differential evolution is then used to refine the concentrations of individual drugs in the input data sets based on feedback from the worm movement data. Effective drug combination is reached in the fourth iteration with minimal centroid velocity (<10 μm/s) and large track curvature (>50 mm$^{-1}$). The winning combination is composed of 1 μM levamisole, 4 μM tribendimidine, 12.5 μM pyrantel, and 100 μM methyridine. Much reduced movement ability is achieved by the winning combination consisting of the four anthelmintics at dosage levels lower than their individual $EC_{50}$ values. The winning combination exhibits high potency as achieved by the maximal dosage of the four drugs but with reduced toxicity and lower cost of drug consumption. The FSC-guided drug combinatorial therapy on a small-animal model discussed here shows that a purely engineering approach for multidrug resistance in nematodes is possible with relevance to other model organisms and parasites.

## MATERIALS AND METHODS
### Microfluidic chip design
The microfluidic device has been previously used to measure the toxicity of *C. elegans* to liquid cyanide solutions (*18*). Briefly, the chip was designed in AutoCAD to have three identical microfluidic chambers (Fig. 2). The chambers were capsule-shaped and without any sharp edges to allow for the smooth movement of *C. elegans*. The input port was tapered leading into the device to permit easy injection of worms into the chamber but prevent the exit of worms. Three chambers can be accommodated within the microscope's field of view and imaged simultaneously. The chip design was sent to an outside vendor (FineLine Imaging) for printing the physical masks. After the masks were obtained, an ultraviolet-sensitive polymer, SU-8, was spin-coated on a 3-inch silicon wafer to a thickness of about 80 μm. The patterns on the physical mask were transferred onto the SU-8, which is then developed. A polydimethylsiloxane (PDMS) polymer was put on the developed SU-8 layer, allowed to dry at room temperature, and later peeled off the SU-8 layer. Subsequently, access holes were punched in the PDMS layer, which is irreversibly bonded to a glass slide to create the microfluidic chip.

### Imaging setup and real-time worm tracking

Postfabrication, the chip was secured on the stage of a Leica MZ16 transmission stereozoom microscope using a plastic tape. Individual chambers were filled with M9 buffer or drug solution using a Tygon microbore tubing (inner diameter, 0.51 mm; outer diameter, 1.52 mm) connected to a 5-ml syringe. Single L4-stage *C. elegans* were picked using a sterilized worm picker and dropped on the input ports of the microfluidic chambers. By applying a small amount of pressure through the syringe, individual worms entered the chamber and start swimming in the liquid medium. All experiments were performed at 20° to 22°C using L4-stage worms. Earlier, wild-type *C. elegans* were grown on nutrient growth medium agarose plates with *Escherichia coli* OP50 bacterial food. The microscope was connected to a high-speed QICAM 12-bit Mono Fast 1394 cooled digital camera with QCapture Pro software, which allows us to record worm behavior across three microfluidic chambers at real time. Digital images (1392 × 1040 pixels) were captured from the recording system for 600 s at one frame per second and were later appended and stitched into single or multiple AVI files for further analyses.

A custom worm tracking program was written in C++ programming language. The program reads the input video file, segments the video into a series of images, identifies the worm body from the background, and then records the *x-y* coordinates of the body's centroid through all the images of the video. The program was able to track worms moving forward or backward, flailing, paused intermittently or indefinitely, and touching other worms or edges of microfluidic devices. In addition, the program adequately keeps track of the worm's boundary throughout the video as it freely moves within the microfluidic chamber, which is often difficult if the worm collides with or touches with other objects in the vicinity. The tracked centroid velocities of individual worms were outputted into an Excel workbook in the form of lists of *x-y* coordinates as a function of time. Statistical analysis and dose-response curve fitting of the generated data were performed using GraphPad Prism. In this software, the steps involve entering the data points, transforming the *x* axis values to log scale, and normalizing the *y* axis values. Then, the "dose response" option was selected followed by the "nonlinear regression" for curve fitting. The prism provides options for the available types of nonlinear regression fits, and the "least square fit" is chosen. The centroid data were used to calculate the instantaneous velocity and average velocity of the worms. In addition, we estimated the average curvature of the worm tracks from the different experiments using the Menger curvature formulation (*58*).

Using microfluidics and automated worm tracking software, the average centroid velocity and track curvature were suitable parameters to score drug-induced inactivity in *C. elegans*. The signatures of the track path (such as velocity, distance moved, and track curvature) correlate with the degradation of the worm's physical ability (*54*–*57*). Because the track paths were recorded at a relatively lower resolution and frame rate, it was possible to observe the behavior of multiple worms within the field of view, which increases the throughput. In comparison, other health metrics related to the body posture (such as thrashing rate, bending frequency, amplitude, and wavelength) were recorded on single worms at a much higher resolution and frame rate, and have low correlation to drug-induced inactivity of *C. elegans* (*54*–*57*). Manual scoring of drug-induced death of *C. elegans* by applying a stimulus to induce motion was feasible on a macroscopic plate or inhibition assays. However, there were definite benefits of using the tracking software over manual scoring of worm death because it provides a means to track the reduced velocity at real time with the advantages of data reproducibility, reduced labor, and elimination of human bias during data collection (*56*, *57*).

### FSC search technique

The FSC search technique adopted in this study was according to earlier publications (*46*–*48*). Figure 1 demonstrates this FSC optimization process flow. First, the dose response of *C. elegans* to the four chosen drugs was conducted. Then, the FSC technique was applied that consists of the following steps: (i) Drug combination pool with four random combinations (P group) were generated and experimentally tested. (ii) Feedback search algorithm named differential evolution (DE) was adopted to generate another four potentially better drug combinations to be tested in the next round of experiments (T group). DE simulates the natural evolution process by applying "mutation" to each drug combination in the P group (low percentage alterations at a few drug doses) and "crossover" between drug combinations within the P group (replacing a portion of one drug combination with another drug combination). The efficacy of each combination in the T group was also experimentally evaluated. The winning drug combinations from the P and T groups were retained to update the P group. The updated P group goes through DE calculation again to generate another T group for comparison. As a result, the P group evolves to be more and more potent (*46*, *47*). (iii) The DE-suggested drug combinations for the four original drug combinations were tested for their antiworm efficacy. The superior ones were used to update the drug combination pool (P group). (iv) Feedback calculation and experimental evaluation described in the steps (i to iii) were repeated until the desired drug combination was identified. The FSC technique was programmed with the MATLAB software. In each iteration, we fed the program with the drug combination's composition and the efficacy of both the P and T groups. The program updates the P group and generates a new T group to be tested in the next iteration.

**Acknowledgments:** We thank J. Powell-Coffman for her insights in the *C. elegans* experiments. **Funding:** This work was partially supported by the State Key Laboratory of Robotics open grant (2016-006 to X.D.) and the China Science and Technology Innovation Zone (17-163-15-XJ-002-002-09 to X.D.). This work was also partially supported by the U.S. National Science Foundation (IDBR-1556370 and CBET-1150867 to S.P.) and the Defense Threat Reduction Agency (HDTRA1-15-1-0053 to S.P.). **Author contributions:** X.D., C.-M.H., and S.P. conceived the original concept. X.D., C.-M.H., Z.N., and S.P. supervised the project. X.D., W.S., and C.-M.H. conducted the algorithmic searches. Z.N. and T.K wrote the tracking software and collected the behavioral data. X.D., Z.N., T.K., W.S., C.-M.H., and S.P. analyzed the data. X.D. and S.P. wrote the manuscript with inputs from all the other authors. All authors discussed the results and contributed to the revisions. **Competing interests:** C.-M.H. is the inventor on a patent related to this work (U.S. patent no. 8,232,095B2, dated 31 July 2012). The authors declare no other competing interests. **Data and materials availability:** All data needed to evaluate the conclusions in the paper are present in the paper. Additional data related to this paper may be requested from the authors.

Submitted 16 June 2017
Accepted 13 September 2017
Published 4 October 2017
10.1126/sciadv.aao1254

**Citation:** X. Ding, Z. Njus, T. Kong, W. Su, C.-M. Ho, S. Pandey, Effective drug combination for *Caenorhabditis elegans* nematodes discovered by output-driven feedback system control technique. *Sci. Adv.* **3**, eaao1254 (2017).

# Science Advances

**Effective drug combination for *Caenorhabditis elegans* nematodes discovered by output-driven feedback system control technique**

Xianting Ding, Zach Njus, Taejoon Kong, Wenqiong Su, Chih-Ming Ho and Santosh Pandey

*Sci Adv* **3** (10), eaao1254.
DOI: 10.1126/sciadv.aao1254